\documentclass[10pt,a4paper,oneside,onecolumn]{article}
\usepackage[a4paper, textheight=700pt, textwidth=500pt]{geometry}
\usepackage[utf8]{inputenc}
\usepackage[T1]{fontenc}
\usepackage{amsmath}
\usepackage{amsfonts}
\usepackage{amssymb}
\usepackage{graphicx}
\usepackage{braket}
\usepackage{color}
\usepackage{float}
\usepackage{bbm}
\usepackage{pdfpages}
\usepackage[labelfont={bf}]{caption}
\usepackage[affil-it]{authblk}
\usepackage[numbers,sort&compress,comma]{natbib}
\usepackage{color} 
\definecolor{rltred}{rgb}{0.75,0,0}  
\definecolor{rltgreen}{rgb}{0,0.5,0}
\definecolor{rltblue}{rgb}{0,0,0.75}
\usepackage{hyperref}
\hypersetup{colorlinks,%
	hypertexnames=true,%
	pdfview=FitH,%
	pdfstartview=FitH,%
	linkcolor=black,%
	citecolor=black,%
	linktocpage=true,%
	pdfproducer={}%
}

\newcommand{\absq}[1]{\left| #1 \right|^2}
\newcommand{\de}[2]{\frac{\mathrm{d}#1}{\mathrm{d}#2}}

\newcommand{\im}{\mathrm{i}}

\newcommand{\beginsupplement}{%
	\setcounter{table}{0}
	\renewcommand{\thetable}{S\arabic{table}}%
	\setcounter{figure}{0}
	\renewcommand{\thefigure}{S\arabic{figure}}%
}

\author[1]{M. Horodynski}
\author[1]{M. K\"uhmayer}
\author[1]{A. Brandst\"otter}
\author[1]{K. Pichler}
\author[2]{Y. V. Fyodorov}
\author[3]{U. Kuhl}
\author[1,*]{S. Rotter}

\affil[1]{Institute for Theoretical Physics, Vienna University of Technology (TU Wien), A-1040 Vienna, Austria}
\affil[2]{Department of Mathematics, King's College London, London WC2R 2LS, UK}
\affil[3]{Universit\'{e} C\^{o}te d'Azur, CNRS, LPMC, 06108 Nice, France}
\affil[*]{Corresponding author: stefan.rotter@tuwien.ac.at}

\title{Optimal Wave Fields for Micro-manipulation \\in Complex Scattering Environments: supplementary material}
\date{}

\begin{document}
	


# Optimal Wave Fields for Micro-manipulation in Complex Scattering Environments

Michael Horodynski[1], Matthias Kühmayer[1], Andre Brandstötter[1], Kevin Pichler[1],

Yan V. Fyodorov[2], Ulrich Kuhl[3], and Stefan Rotter[1]

**Affiliations:**

[1] Institute for Theoretical Physics, Vienna University of Technology (TU Wien), A-1040 Vienna, Austria

[2] Department of Mathematics, King's College London, London WC2R 2LS, United Kingdom

[3] Institut de Physique de Nice, Université Côte d'Azur, CNRS, F-06100 Nice, France

**The manipulation of small objects with light has become an indispensable tool in many areas of research ranging from physics to biology and medicine[1–7]. Here we demonstrate how to implement micro-manipulation at the optimal level of efficiency for targets of arbitrary shape and inside complex environments such as disordered media. Our approach is to design wave-fronts in the far-field[8–15] that have optimal properties in the near-field of the target such as to apply to it the strongest possible force, pressure or torque as well as to achieve the most efficient focus at the target position. Free of any iterative optimization, our approach only relies on a simple eigenvalue problem established from the scattering matrix of the system and its dependence on the target parameters. To illustrate this theoretical concept, we perform a proof-of-principle experiment in the microwave regime, which fully confirms our predictions.**

The 2018 Nobel prize for physics was, in part, awarded to Arthur Ashkin for his seminal work on optical tweezers[1], which revolutionised the way one can trap and manipulate

1

nanoscale dielectric particles remotely using only a laser beam. The key idea is that the field pattern of the beam creates forces at the position of the target particle that can be used for very different purposes, like for cellular manipulation[2,3], fluid dynamics[4], micro-robotics[5] and for tests of fundamental physics[6,7].

A remaining and critical challenge, especially for applications in biology and soft-matter, stems from the fact that targets are often embedded within complex media like tissues, which scatter the incoming beam and thus destroy the field pattern necessary to trap or manipulate a target efficiently[16]. Even when guiding light to a target directly through a multi-mode fibre one faces similar problems since any imperfections in the fibre also lead to non-trivial field distortions[17].

To remedy the difficulties posed by unwanted scattering processes, the emerging field of wave front shaping[8–10,12–15] has developed a new set of tools that are meanwhile also being considered for versatile micro-manipulation strategies far beyond the standard particle trap of a Gaussian beam[18]. In a first generation of studies, iterative computational optimization schemes were used to increase the stiffness of optical traps[19,20], to enhance micromanipulation capabilities[21], or to trap nano-particles across a highly turbid medium[16]. Yet, iterative approaches genuinely suffer from the deficiency that they may not reach a global optimum or that they may require a prohibitively large number of computational steps. A series of works also circumvent the need for an iteration to find the optimal state by recognizing the power of quadratic forms[22] to optimize quantities like the transferred power, momentum and torque[22–24] or the stiffness of an optical trap[25]. While being very efficient, setting up the quadratic form itself requires the knowledge of the full Green's function between the asymptotic far-field and the near-field of the target[22]. For targets inside a disordered environment this knowledge is typically not available, however. Only for the specific case when the

2

target is a single scatterer, which is individually addressable from the far-field, it was recently shown that one may work instead with the reduced information stored in the asymptotic scattering matrix elements[23,24].

Here we overcome these limitations by introducing a broadly applicable framework that achieves micromanipulation of targets that can be very complex either in terms of their own shape or in terms of the aberrating (inhomogeneous) environment surrounding them. Our procedure operates at the optimal efficiency level through the resolution of a simple linear eigenvalue problem (without iteration). Specifically, we provide a straightforward way to determine, out of all possible states with the same incoming flux in the far-field, the optimal state with the unique near-field profile that applies the highest possible force, pressure or torque on a non-absorbing target.

Our starting point is the time-delay operator $Q$ introduced by Wigner and Smith[26,27] based on a system's scattering matrix $S$ that relates all incoming with all outgoing field states, $S|\psi^{\text{in}}\rangle = |\psi^{\text{out}}\rangle$. Involving the derivative of the $S$-matrix with respect to the frequency $\omega$ of the incoming wave, the time-delay operator $Q = -iS^{-1}dS/d\omega$ has eigenstates (also known as "principal modes") with a well-defined time-delay between entering and exiting a given system[15,28]. This property results in the remarkable feature that the output profile of these unique states stays unchanged for small variations of the frequency[15,29–35]. A more general class of Wigner-Smith operators was recently introduced[36] by generalizing the time-delay operator $Q$ to feature a derivative of the scattering matrix $S$ with respect to an arbitrary parameter $\alpha$, rather than to the frequency $\omega$. These generalised Wigner-Smith (GWS) operators $Q_\alpha$ possess eigenvectors $|u_\alpha^i\rangle$ and eigenvalues $\theta_\alpha^i$ that fulfil

$$Q_\alpha |u_\alpha^i\rangle = -\mathrm{i} S^{-1} \frac{\mathrm{d}S}{\mathrm{d}\alpha} |u_\alpha^i\rangle = \theta_\alpha^i |u_\alpha^i\rangle. \tag{1}$$

For flux-conserving systems, the scattering matrix is unitary and the GWS operators $Q_\alpha$ are Hermitian. To give the orthogonal eigenvectors $|u_\alpha^i\rangle$ and the real eigenvalues $\theta_\alpha$ of $Q_\alpha$ a physical meaning, one can proceed by analogy: the eigenvalues of the conventional Wigner-Smith time-delay operator $Q$ (involving a derivative by frequency $\omega$) correspond to the so-called "proper delay times". Since time and frequency are conjugate quantities (in the sense used for a Fourier transform), the generalised Wigner-Smith operator $Q_\alpha$ (involving a derivative by $\alpha$) should have eigenvalues that represent the conjugate quantity to $\alpha$ (if $\alpha$ is, e.g., a position then the eigenvalues of $Q_\alpha$ would be momenta). Most importantly, however, the quantity $\alpha$ can not only be a global parameter that describes the entire scattering system, but also a local parameter such as the position of a single scatterer inside a disordered medium[36]. For this specific case, it was found[36] that the eigenstates $|u_\alpha^i\rangle$ of $Q_\alpha$ transfer onto this scatterer a momentum that was numerically observed to be linearly proportional to the corresponding eigenvalue $\theta_\alpha^i$. Here we show that this observation is not just of empirical nature, but that one can, in fact, formulate fundamental relationships between all the central observables in micro-manipulation like the transfer of momentum, pressure, torque etc. onto a target and the eigenvalues of $Q_\alpha$ when $\alpha$ is chosen to be the target's position, radius or angular orientation.

To arrive at these analytical relations, we consider a dielectric landscape $\varepsilon(\vec{r})$ as described by the scalar Helmholtz equation $[\Delta + k^2 \varepsilon(\vec{r})]\psi = 0$. Here, $k = |\vec{k}|$ is the wavenumber, $\psi$ is the transverse component of the electric field, $\Delta$ is the Laplacian in two dimensions, $\varepsilon(\vec{r})$ the spatially varying dielectric function and $\vec{r}$ is the position vector. Using the framework of mesoscopic transport theory[15,37], we link the scattering

4

landscape as determined by $U(\vec{r}) = k^2 \varepsilon(\vec{r})$ with the asymptotic scattering amplitudes as contained in the $S$-matrix (see supplemental material). Describing any conceivable variation of $U(\vec{r})$ by a derivative with respect to some arbitrary parameter $\alpha$ (local or global), results in the following basic and compact relation,

$$\langle \chi | Q_\alpha | \chi \rangle = \left\langle \psi_\chi \left| \frac{\mathrm{d}U}{\mathrm{d}\alpha} \right| \psi_\chi \right\rangle / 2, \tag{2}$$

in which the GWS operator $Q_\alpha$ emerges naturally. Here, $|\chi\rangle$ stands for an arbitrary asymptotic wave state impinging on the system (typically a vector of modal amplitudes), whereas $|\psi_\chi\rangle$ is the resulting wave field inside the scattering system (typically a spatially resolved field distribution). Equation (2) can be exploited in multiple ways: given, e.g., a certain scattering wave function $|\psi_\chi\rangle$ one can modify the system (via $\mathrm{d}U/\mathrm{d}\alpha$) such as to achieve a desired change in the scattering matrix $S$ (via $Q_\alpha$). Here, we consider the opposite scenario and ask how to shape the incoming wave front $|\chi\rangle$ to manipulate a target in an optimal way. The type of manipulation we aim to perform is encapsulated in the variable $\alpha$, which the potential $U$ depends on parametrically. Specifically, we will demonstrate how specific changes $\mathrm{d}U/\mathrm{d}\alpha$ lead to all the essential quantities in micromanipulation: a small rotation of the target as parametrized by the angle $\alpha = \varphi$ yields the torque transferred onto it; shifting the target $\alpha = \vec{x}$ provides the momentum transfer; varying the radius $\alpha = R$ of a circular target grants control over the radiation pressure exerted on it and changing the value of the dielectric constant $\alpha = \varepsilon$ determines the wave intensity inside the target (which is equivalent to the dwell time inside the target's volume).

We now introduce all of these individual micromanipulation tools explicitly and verify them both numerically and experimentally using a rectangular multimode waveguide. The employed microwave setup is especially suited to give access both to the full

5

scattering matrix as well as to the internal field distribution (see Methods). Additionally, any desired wave front can be injected to the system in situ and variations of the geometry can be performed with subwavelength precision. The platform is made out of aluminium in which we place randomly distributed Teflon scatterers around a target in the middle to simulate a disordered medium (Fig. 1). Waves are injected from both sides at a frequency of 12.75 GHz, resulting in a wavelength of $\lambda = 2\pi/k = 2.35$ cm where 8 transversal modes can propagate.

In the first example we determine the parameter $\alpha$ to be the orientation angle $\varphi$ of a complex-shaped metallic target embedded in an empty waveguide (Fig. 2a) and of a metallic square target inside a disordered medium (Fig. 2c). The corresponding GWS-operator, $Q_\varphi = -iS^{-1}dS/d\varphi$, involves measuring the differential response of the scattering matrix $S$ with respect to changes of the target's orientation angle $\varphi$ (to evaluate the derivative $dS/d\varphi$ we measure $S$ for $\varphi = \pm 5°$ and $0°$). As we prove in the supplemental material, the eigenvalues $\theta_\varphi^i$ of this operator $Q_\varphi$ are exactly one half of the torque $M_z$ exerted on the target,

$$M_z(|u_\varphi^i\rangle) \equiv \int_\mathcal{C} [\vec{m}_\perp(\vec{c}) \times \vec{n}(\vec{c}) |\partial_{\vec{n}} \psi(\vec{c})|^2]_z ds = 2\theta_\varphi^i, \qquad (3)$$

where the integral is taken along the target's boundary described by the curve $\mathcal{C}$ and parametrised by $\vec{c}$. The expression $\vec{n}(\vec{c})|\partial_{\vec{n}}\psi(\vec{c})|^2$ denotes the normal force at every point of the boundary exercised by the electric field and $\vec{m}_\perp(\vec{c})$ is the lever (the part of the distance from the boundary to the target's center of mass that is normal to $\vec{n}$). We emphasise that Eq. 3 has no free parameters and is exact in the sense that no approximations are involved in deriving it. Particularly appealing for micro-manipulation is now the feature that Eq. 3 directly connects the maximally achievable torque to the largest eigenvalue $\theta_\varphi^{\max}$, which, in turn, is obtained by injecting the corresponding

6

eigenvector $|u_\varphi^{\max}\rangle$ into the system. The experimentally measured intensity distribution of the eigenstates of $Q_\varphi$ with the largest eigenvalues can be seen in Figs. 2a and 2c for the two configurations of a complex target without disorder and for a simple target inside the disorder. Both figures confirm our predictions by way of the visible focus of the field onto the target's corners, where the lever is maximal and thereby also the corresponding torque.

In the supplemental material we extend Eq. 3 to the case of a dielectric target and provide equivalent relations with links between the spatial displacement of the target and the momentum transferred onto it as well as between the radius of a circular scatterer and the radiation pressure acting on it. We also investigate there the common situation that only a sub-unitary part of the scattering matrix is available for the evaluation of the GWS-operator. Our results show that there is still an excellent agreement between the eigenvalues $\theta_\varphi^i$ and the transferred torque $M_z$ as long as the available part of the scattering matrix contains sufficient information on the target.

One aspect one observes in Fig. 2a is that the displayed eigenstate of $Q_\varphi$ exerts not only a torque, but also a force that moves the target in a certain direction. If one is now interested in rotating the target while keeping the position of its center of mass unchanged, this can be achieved by choosing two eigenstates $|u_\varphi^1\rangle$ and $|u_\varphi^2\rangle$ of $Q_\varphi$ that both apply a high torque to the target in the same direction, but whose linear momentum transfer has opposite signs. Out of these two states we construct a superposition, $|s\rangle = |u_\varphi^1\rangle + |u_\varphi^2\rangle$, for which the linear momentum transfer disappears, as illustrated by the resulting spatial intensity distribution of the state $|s\rangle$ in Fig. 2b, which focuses onto both of the opposite corners of the target.

7

A last property of the target that we have so far not exploited is its dielectric constant $\varepsilon$. We find that a derivative with respect to $\varepsilon$ in a corresponding GWS operator yields the integrated intensity $I$ of the wave field inside the target (in analogy to the local change in the potential yielding the local density of states in quantum scattering[38,39]). Specifically, we derive a linear relation between the eigenvalue $\theta_\varepsilon^i$ of $Q_\varepsilon$ and the integrated wave intensity of the scattering state $\psi$ injected through the eigenvector $|u_\varepsilon^i\rangle$ (see supplement),

$$I(|u_\varepsilon^i\rangle) \equiv \int |\psi|^2 dA = 2\theta_\varepsilon^i/k^2 . \qquad (4)$$

The integral domain $A$ is the area of our target scatterer whose dielectric constant is changed in the differentiation of $Q_\varepsilon = -iS^{-1}\, dS/d\varepsilon$ and $\psi$ is the electric field distribution of the corresponding eigenstate $|u_\varepsilon^i\rangle$. Maximizing a wave's intensity inside a chosen target thus involves the injection of the eigenstate $|u_\varepsilon^{\max}\rangle$ associated to the largest eigenvalue $\theta_\varepsilon^{\max}$. In the microwave experiments we implemented an effective change of the dielectric constant ($\varepsilon_r = 1.4884, 1.7689$ and $2.0736$) by varying the height ($h = 4, 6$ and $8$ mm) of the target Teflon cylinder with radius $R = 6$ mm. The spatial intensity distribution of the state $|u_\varepsilon^{\max}\rangle$ corresponding to the maximal eigenvalue in the experiment is shown in Fig. 3a. One can clearly see that the scattering state has a strong intensity build-up inside the chosen target. To prove that the focus we achieve in this way is, indeed, optimal, we successfully compare it numerically to an independent technique[40], which requires access to the Green's function at the target and is thus much more costly to implement experimentally (see supplemental material).

We also highlight here that the integrated intensity inside a target, which was shown to be proportional to the eigenvalue $\theta_\varepsilon$ in Eq. 4, is itself proportional to the dwell-time of

8

the wave state $\psi$ inside the target[15,28]. In this way, the GWS operator $Q_\varepsilon$ allows us to measure the dwell-time not only inside the entire scattering region (as with the conventional Wigner-Smith operator $Q_\omega$), but also in any desired sub-part of it.

For a realization of our approach in the optical regime, we envision a setup similar to the one used in the first successful experiment on trapping behind a turbid medium[16]. There, the trapped particles were about 1 μm in size and held in place by approximately 1 mW of laser power at the target position. The important ingredient to be added now is to shift the target particle in its position, angle, size or in its refractive index such as to produce a noticeable trace in the measured scattering matrix entering the GWS operator. For this purpose we suggest to employ beams of ultrasound such as those already in use to produce very localized changes in the dielectric constant[10,13,41,42] or to take advantage of the information stored in the acousto-optic transmission matrix[42]. Alternatively, one could also work with the kinetic or self-propelled movement of a target in an otherwise static background, such as employed already in advanced focusing techniques[43,44]. A definite advantage of our approach is that it also works for the realistic case that parts of the scattering matrix are inaccessible experimentally (see supplemental material for details).

To conclude, we present a general framework for optimal micromanipulation with targets of arbitrary shape and in arbitrarily complex environments. We successfully tested this concept experimentally and envision it to be a key for breaking the barrier imposed by disordered media on the applicability of optical tools for manipulating objects inside of them. Ultimately, our work may serve as a guidepost towards a new generation of micro-manipulation experiments with wave-front shaping protocols that continuously operate at the optimal level based on a real-time monitoring of a system's scattering matrix.

**References**

1. Ashkin, A., Dziedzic, J. M., Bjorkholm, J. E. & Chu, S. Observation of a single-beam gradient force optical trap for dielectric particles. *Opt. Lett.* **11**, 288 (1986).

2. Thalhammer, G., Steiger, R., Bernet, S. & Ritsch-Marte, M. Optical macro-tweezers: trapping of highly motile micro-organisms. *J. Opt.* **13**, 044024 (2011).

3. Bowman, R. *et al.* Position clamping in a holographic counterpropagating optical trap. *Opt. Express* **19**, 9908 (2011).

4. Jannasch, A., Mahamdeh, M. & Schäffer, E. Inertial Effects of a Small Brownian Particle Cause a Colored Power Spectral Density of Thermal Noise. *Phys. Rev. Lett.* **107**, (2011).

5. Palima, D. & Glückstad, J. Gearing up for optical microrobotics: micromanipulation and actuation of synthetic microstructures by optical forces: Gearing up for optical microrobotics. *Laser Photonics Rev.* **7**, 478–494 (2013).

6. Kheifets, S., Simha, A., Melin, K., Li, T. & Raizen, M. G. Observation of Brownian Motion in Liquids at Short Times: Instantaneous Velocity and Memory Loss. *Science* **343**, 1493–1496 (2014).

7. Li, T. Millikelvin Cooling of an Optically Trapped Microsphere in Vacuum. in *Fundamental Tests of Physics with Optically Trapped*

*Microspheres* 81–110 (Springer New York, 2013). doi:10.1007/978-1-4614-6031-2_6

8. Vellekoop, I. M. & Mosk, A. P. Focusing coherent light through opaque strongly scattering media. *Opt. Lett.* **32**, 2309 (2007).

9. van Putten, E. G. *et al.* Scattering Lens Resolves Sub-100 nm Structures with Visible Light. *Phys. Rev. Lett.* **106**, (2011).

10. Xu, X., Liu, H. & Wang, L. V. Time-reversed ultrasonically encoded optical focusing into scattering media. *Nat. Photonics* **5**, 154–157 (2011).

11. Dholakia, K. & Čižmár, T. Shaping the future of manipulation. *Nat. Photonics* **5**, 335–342 (2011).

12. Mosk, A. P., Lagendijk, A., Lerosey, G. & Fink, M. Controlling waves in space and time for imaging and focusing in complex media. *Nat. Photonics* **6**, 283–292 (2012).

13. Judkewitz, B., Wang, Y. M., Horstmeyer, R., Mathy, A. & Yang, C. Speckle-scale focusing in the diffusive regime with time reversal of variance-encoded light (TROVE). *Nat. Photonics* **7**, 300–305 (2013).

14. Horstmeyer, R., Ruan, H. & Yang, C. Guidestar-assisted wavefront-shaping methods for focusing light into biological tissue. *Nat. Photonics* **9**, 563–571 (2015).

15. Rotter, S. & Gigan, S. Light fields in complex media: Mesoscopic scattering meets wave control. *Rev. Mod. Phys.* **89**, (2017).

16. Čižmár, T., Mazilu, M. & Dholakia, K. In situ wavefront correction and its application to micromanipulation. *Nat. Photonics* **4**, 388–394 (2010).

17. Richardson, D. J., Fini, J. M. & Nelson, L. E. Space-division multiplexing in optical fibres. *Nat. Photonics* **7**, 354–362 (2013).

18. Dholakia, K. & Čižmár, T. Shaping the future of manipulation. *Nat. Photonics* **5**, 335–342 (2011).

19. Polin, M., Ladavac, K., Lee, S.-H., Roichman, Y. & Grier, D. G. Optimized holographic optical traps. *Opt. Express* **13**, 5831 (2005).

20. Taylor, M. A., Waleed, M., Stilgoe, A. B., Rubinsztein-Dunlop, H. & Bowen, W. P. Enhanced optical trapping via structured scattering. *Nat. Photonics* **9**, 669–673 (2015).

21. Lee, Y. E., Miller, O. D., Homer Reid, M. T., Johnson, S. G. & Fang, N. X. Computational inverse design of non-intuitive illumination patterns to maximize optical force or torque. *Opt. Express* **25**, 6757 (2017).

22. Mazilu, M., Baumgartl, J., Kosmeier, S. & Dholakia, K. Optical Eigenmodes; exploiting the quadratic nature of the energy flux and of scattering interactions. *Opt. Express* **19**, 933–945 (2011).

23. Fernandez-Corbaton, I. & Rockstuhl, C. Unified theory to describe and engineer conservation laws in light-matter interactions. *Phys. Rev. A* **95**, (2017).

24. Liu, Y. *et al.* Optimal Nanoparticle Forces, Torques, and Illumination Fields. *ACS Photonics* (2019). doi:10.1021/acsphotonics.8b01263

25. Taylor, M. A. Optimizing phase to enhance optical trap stiffness. *Sci. Rep.* **7**, (2017).

26. Wigner, E. P. Lower Limit for the Energy Derivative of the Scattering Phase Shift. *Phys. Rev.* **98**, 145–147 (1955).

27. Smith, F. T. Lifetime Matrix in Collision Theory. *Phys. Rev.* **118**, 349–356 (1960).

28. Winful, H. G. Delay Time and the Hartman Effect in Quantum Tunneling. *Phys. Rev. Lett.* **91**, 260401 (2003).

29. Fan, S. & Kahn, J. M. Principal modes in multimode waveguides. *Opt. Lett.* **30**, 135 (2005).

30. Rotter, S., Ambichl, P. & Libisch, F. Generating Particlelike Scattering States in Wave Transport. *Phys. Rev. Lett.* **106**, (2011).

31. Carpenter, J., Eggleton, B. J. & Schröder, J. Observation of Eisenbud–Wigner–Smith states as principal modes in multimode fibre. *Nat. Photonics* **9**, 751–757 (2015).

32. Xiong, W. *et al.* Spatiotemporal Control of Light Transmission through a Multimode Fiber with Strong Mode Coupling. *Phys. Rev. Lett.* **117**, (2016).

33. Gérardin, B. *et al.* Particlelike wave packets in complex scattering systems. *Phys. Rev. B* **94**, (2016).

34. Böhm, J., Brandstötter, A., Ambichl, P., Rotter, S. & Kuhl, U. In situ realization of particlelike scattering states in a microwave cavity. *Phys. Rev. A* **97**, 021801 (2018).

35. Brandstötter, A., Girschik, A., Ambichl, P. & Rotter, S. Shaping the Branched Flow of Light through Disordered Media. *ArXiv190405229 Cond-Mat Physicsnlin Physicsphysics* (2019).

36. Ambichl, P. *et al.* Focusing inside Disordered Media with the Generalized Wigner-Smith Operator. *Phys. Rev. Lett.* **119**, (2017).

37. Fyodorov, Y. V. & Sommers, H.-J. Statistics of resonance poles, phase shifts and time delays in quantum chaotic scattering: Random matrix approach for systems with broken time-reversal invariance. *J. Math. Phys.* **38**, 1918–1981 (1997).

38. Brandbyge, M. & Tsukada, M. Local density of states from transmission amplitudes in multichannel systems. *Phys. Rev. B* **57**, R15088–R15091 (1998).

39. Davy, M., Shi, Z., Wang, J., Cheng, X. & Genack, A. Z. Transmission Eigenchannels and the Densities of States of Random Media. *Phys. Rev. Lett.* **114**, (2015).

40. Cheng, X. & Genack, A. Z. Focusing and energy deposition inside random media. *Opt. Lett.* **39**, 6324 (2014).

41. Tay, J. W., Lai, P., Suzuki, Y. & Wang, L. V. Ultrasonically encoded wavefront shaping for focusing into random media. *Sci. Rep.* **4**, (2015).

42. Katz, O., Ramaz, F., Gigan, S. & Fink, M. Controlling light in complex media beyond the acoustic diffraction-limit using the acousto-optic transmission matrix. *Nat. Commun.* **10**, 717 (2019).

43. Ma, C., Xu, X., Liu, Y. & Wang, L. V. Time-reversed adapted-perturbation (TRAP) optical focusing onto dynamic objects inside scattering media. *Nat. Photonics* **8**, 931–936 (2014).

44. Zhou, E. H., Ruan, H., Yang, C. & Judkewitz, B. Focusing on moving targets through scattering samples. *Optica* **1**, 227 (2014).

**Supplementary Information** …



# Figures

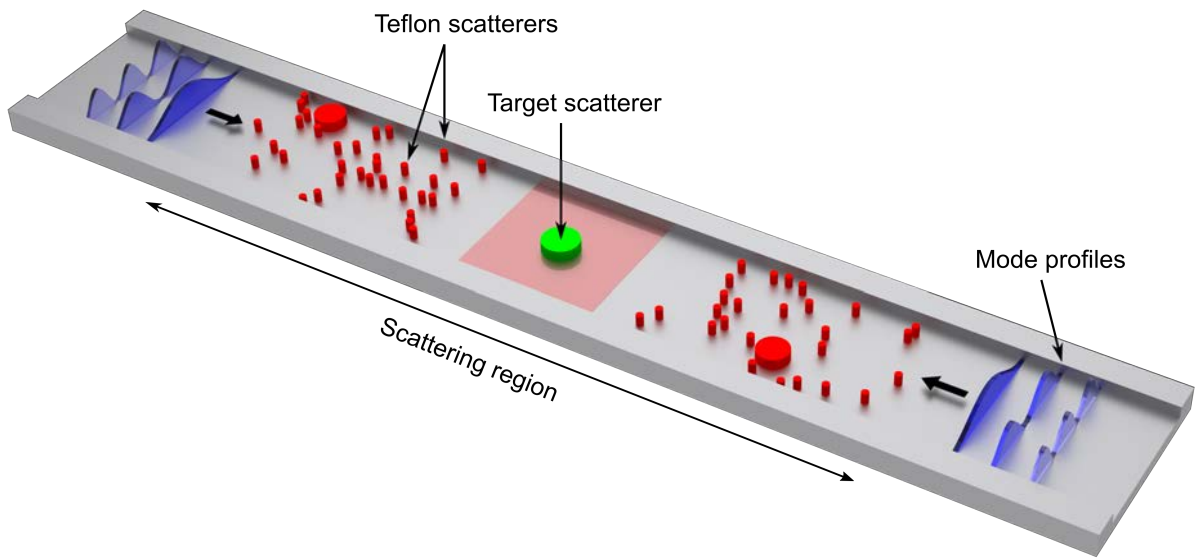

**Figure 1: Experimental setup.** Sketch of the scattering system used in the experiment, consisting of a two-dimensional waveguide geometry (length $L = 60$ cm and width $W = 10$ cm) with hard walls in the transverse direction and leads on the left and right. The top plate (not shown) features a grid of small holes through which the field intensity in the waveguide is measured. Red cylinders indicate the randomly distributed circular Teflon scatterers (dielectric constant $\varepsilon_r = 2.0736$ and radii 2.55 mm or 11 mm) and the green cylinder in the middle indicates a target scatterer, whose material, size and shape is chosen specifically for each GWS-operator we investigate. The surrounding disorder (without the target scatterer) is characterised by a transport mean free path of $l_t = 35$ cm and a scattering mean free path of $l_s = 27$ cm. The red-shaded area indicates the region shown in Figs. 2, 3. Eight antennas on each side (not shown) are used to measure the full scattering matrix $S$ and to inject the desired scattering states.

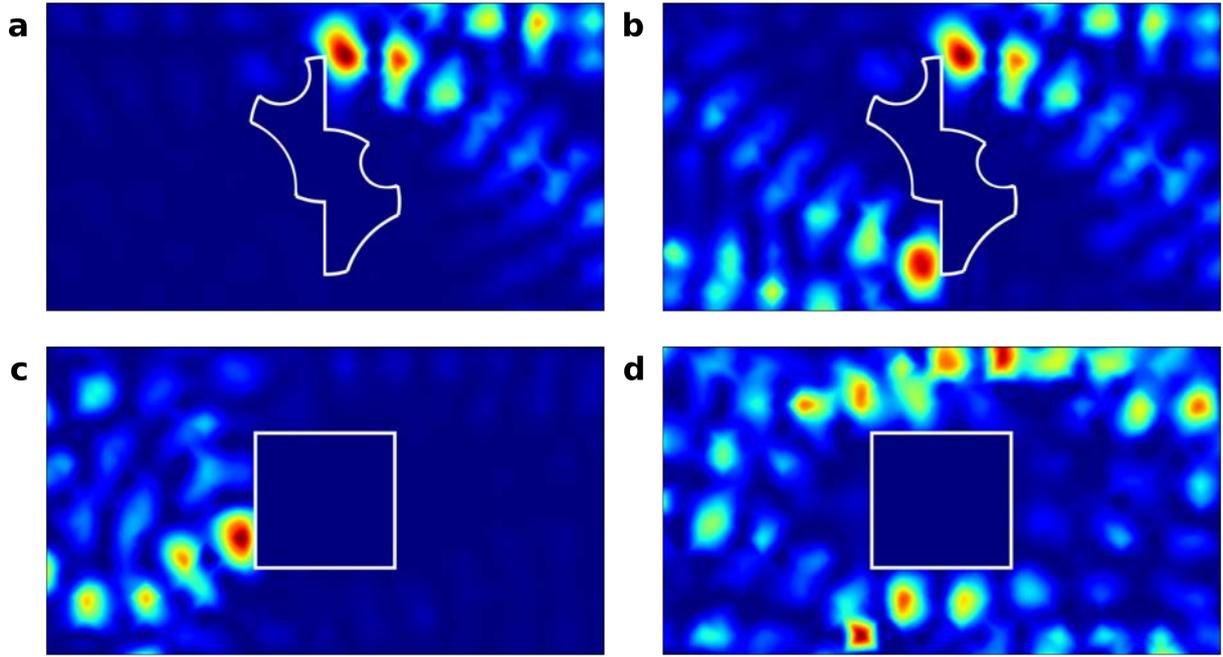

**Figure 2: Optimal transfer of torque.** Experimental measurements of the electric field intensity (colour) for wave states with optimal transfer of torque on a metallic target (white boundary). An asymmetric target of complex shape in an empty waveguide (**a,b**) and a quadratic target in the disordered waveguide (**c,d**) are considered. (The region displayed here corresponds to the red-shaded area in Fig. 1.) **a,** Intensity of the $Q_\varphi$-eigenstate $|u_\varphi^{\max}\rangle$ corresponding to the largest eigenvalue $\theta_\varphi^{\max}$, displaying a focus on the point of maximal lever. **b,** Intensity for a superposition of the eigenstates associated to the two largest eigenvalues, that transfers only torque, but no linear momentum on the target. The corresponding intensity builds up on both of the most protruding parts of the target. For a metallic square target inside the disordered waveguide we compare (**c**) the $Q_\varphi$-eigenstate with the maximal eigenvalue $|\theta_\varphi^{\max}| \approx 5.85$ with (**d**) an eigenstate having a small eigenvalue $|\theta_\varphi^{\text{small}}| \approx 0.2$. In all figures (also below) the colour scale has been adjusted to match the maximum intensity (shown in dark red).

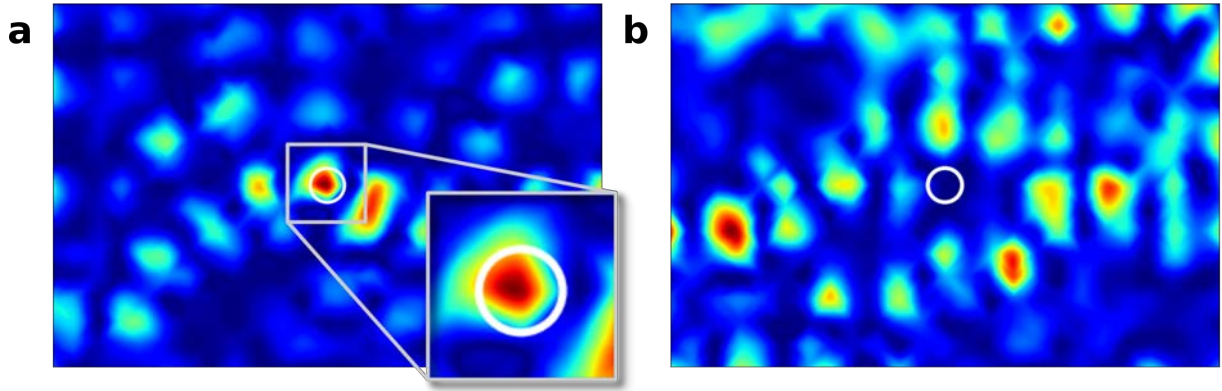

**Figure 3: Focusing with the GWS-operator.** Experimentally measured spatial intensity distribution of states with a well-controlled focus at a target embedded inside a disordered environment (as shown in Fig. 1). We use a dielectric target (white circle) and compare (**a**) the intensity of the $Q_\varepsilon$-eigenstate $|u_\varepsilon^{\max}\rangle$ featuring the largest eigenvalue $|\theta_\varepsilon^{\max}| \approx 1.96$ with (**b**) an eigenstate with a small eigenvalue $|\theta_\varepsilon^{\text{small}}| \approx 0.46$. A strong focus within the target's boundary is observed in **a** (see the inset for a zoom), whereas no focus is observed in **b**.

## Methods

**Experimental setup.** To implement the GWS-operator we use a rectangular aluminium waveguide that is 10 cm wide ($y$ direction), 8 mm high ($z$ direction) and 238 cm long ($x$ direction). Our experiment operates at a frequency of 12.75 GHz, resulting in 8 propagating modes that are fully controlled by 8 antennas placed on each end of the waveguide and reaching about 4 mm into the waveguide resulting in weak coupling. The input states are generated by a vector network analyser (VNA) and get sent to the antennas via altogether 16 IQ modulators, that allow us to control the amplitudes and relative phases. The open ends of the waveguide are plugged by absorbers to avoid back-reflection of waves (to mimic semi-infinite leads). The scattering elements (Teflon cylinders of radii 2.55 mm and 11 mm) are of the same height as the waveguide (8 mm) and concentrated in the middle of the waveguide on a length of 50 cm. In the middle of this scattering region we place our targets of different shapes and sizes (see Fig. 1). Using the altogether 16 attached antennas (8 on each side of the waveguide), we first measure the scattering matrix $S$ for different target configurations. Next, we inject the GWS-eigenstates generated based on these $S$-matrices and measure the field inside the scattering region in an area around the target. We do this with a movable probing antenna that we insert into the waveguide through holes drilled into the top plate of the waveguide in the shape of a rectangular grid with a spacing of 5 mm. The effect of the probing antenna on the field is small, since it only sticks 4 mm into the waveguide, but since it needs to be inserted into the waveguide it cannot measure the intensity inside the target or a scatterer.

**Controlling the input state.** The demonstration of the GWS-operator relies on the full control of each input state as achieved by the 8 antennas we work with on each end of the waveguide in which 8 propagating modes are available at the operational

20

frequency. To shape the amplitudes and relative phases of the signal at each input antenna we use IQ modulators whose settings need to be related to the associated incoming waves. To identify this relationship, we first construct an 'IQ modulator'-basis, where each of the modulators corresponds to a basis vector, i.e., one modulator is fully transmitting, while we set the others to maximum damping. In the second step we measure the field along the $y$-direction for two $x$-positions on both sides of the waveguide. We need at least as many measurements in the $y$-direction as we have propagating modes. To reduce noise, we make more measurements than the necessary 8 on each side, namely 19. In the third step we note that the measured field is a superposition of incoming and outgoing modes, hence the need for measuring at two $x$-positions. To access the modes in this basis we use the following linear relations,

$$a_{\mathrm{m},j}(x) = a_{\mathrm{in},j} e^{-ik_{\mathrm{x},j} x} + a_{\mathrm{out},j} e^{ik_{\mathrm{x},j} x} ,$$

$$a_{\mathrm{m},j}(x+dx) = a_{\mathrm{in},j} e^{-ik_{\mathrm{x},j}(x+dx)} + a_{\mathrm{out},j} e^{ik_{\mathrm{x},j}(x+dx)} ,$$

where $j$ is the number of the respective mode and $k_{\mathrm{x},j} = \sqrt{k^2 - k_{\mathrm{y},j}^2}$, where $k = \frac{2\pi \nu}{c}$, $k_{\mathrm{y},j} = \frac{j\pi}{d}$, $\nu$ is the frequency of the ingoing waves, $c$ is the speed of light, and $d$ is the width of the waveguide in $y$ direction. In order to find the correspondence between IQ modulator settings and specific input states one has to perform a basis transformation from the IQ modulator basis to the transverse mode representation.

**Evaluating the scattering matrix.** For the construction of the GWS-operator we need full knowledge of the system's scattering matrix $S$ for three different configurations. These are $S_0$ for the neutral configuration, $S_+$ for the configuration where we have increased $\alpha$ and $S_-$ for the configuration where we have decreased $\alpha$. This allows us to construct the GWS-operator with a symmetric finite difference approximating the

derivative, i.e., $Q_\alpha \approx -\mathrm{i}\, S_0^{-1}(S_+ - S_-)/(2\Delta\alpha)$. The complex $S$-matrix connects the incoming with the outgoing modes (both flux-normalised), $\vec{a}_{\mathrm{out}} = S\, \vec{a}_{\mathrm{in}}$, which describes a set of $N$ linearly independent equations, where $N$ is the total number of open channels (16 in our case). The scattering matrix $S$ has $N^2$ elements, therefore we need $N$ linearly independent sets of coefficients $\vec{a}_{\mathrm{in}}$ and $\vec{a}_{\mathrm{out}}$ to solve for the scattering matrix elements. As outlined in the previous paragraph, we obtain these coefficients by measuring the field between the scattering region and the antenna array in both sides at $N$ positions in the $y$-direction and for two different positions in the $x$-direction, while only one IQ modulator is transmitting.

**Data processing.** In order to reduce noise effects, we treat the data of the transmission between the external antennas and the scanning antenna with a weighted average in the frequency domain. The resulting intensity data are treated with a bicubic interpolation to generate the spatial intensity distributions shown in Figs. 2 and 3. Since the field cannot be measured inside the target we reconstruct the information about the field inside the target from the mode decomposition which we can calculate from the 16 accessible points along the $y$ direction at the $x$ position of the target.

**Numerical simulations.** In our numerical simulations we solve the scalar Helmholtz equation, $[\Delta + k^2 \varepsilon(\vec{r})]\psi(\vec{r}) = 0$, where $\Delta$ is the Laplacian in two dimensions, $\varepsilon(\vec{r})$ is the spatially varying dielectric function, $k = |\vec{k}| = \omega/c$ is the vacuum wavenumber, $\psi(\vec{r})$ is the $z$-component of the electrical field and all spatially varying functions depend on the position $\vec{r} = (x, y)$. Since for most complex refractive index landscapes there is no analytic solution for the Helmholtz equation, we solve it numerically using an advanced finite-element method (NGSolve open source Finite Element Library at https://ngsolve.org/). The scalar Helmholtz equation in two dimensions is sufficient for

describing the three-dimensional experiment, since the waveguide is only 8 mm high and thus features only one transverse mode in $z$-direction. The absorbers plugged at the ends of the waveguides are implemented into the simulation through 'perfectly matched layers' and on the long edges our scattering region is bounded by hard walls.



\maketitle
\beginsupplement

\section{Generalised Wigner-Smith Operators in One Dimension}
\begin{figure}[t!]
	\centering
	\includegraphics[width=\textwidth]{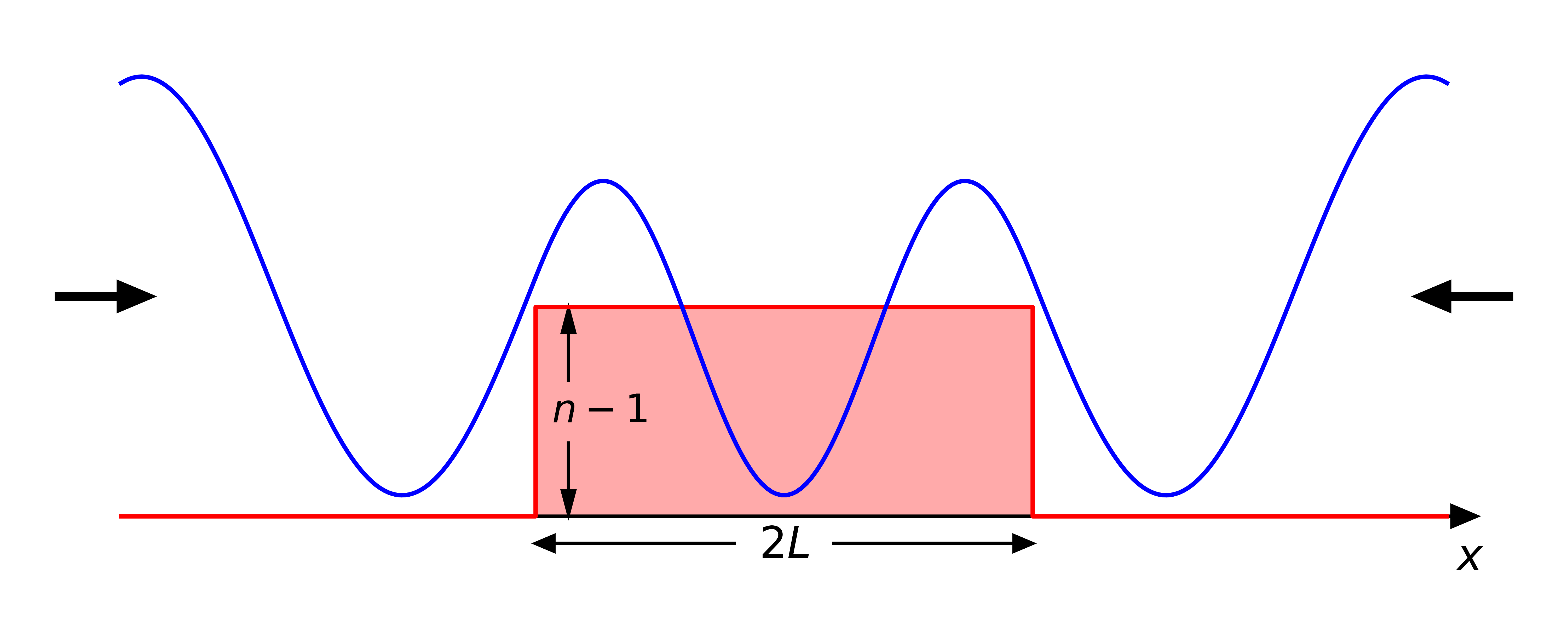}
	\caption{Sketch of the refractive index barrier (red line) of width $2L$ and a constant height of $n-1$. The blue line shows the intensity distribution of the second $Q_n$-eigenstate, $\vec{u}^{2}_{n}$, for a wavelength of $\lambda = 2\pi$, refractive index $n=1.44$ and a length $L=1.5$. The black arrows indicate that plane waves injected from both sides construct the scattering state.}\label{fig:1ds}
\end{figure}

To illustrate how our concept based on the generalised Wigner Smith operator works in a very simple context, we consider here the problem of scattering at a one-dimensional refractive index barrier (see Fig.~\ref{fig:1ds}) of refractive index $n>1$ ranging from $x=-L$ to $x=L$ (the refractive index in the asymptotic regions is $n_0=1$). For this simple problem we can explicitly calculate the scattering matrix $S$ and all the possible GWS-operators $Q_\alpha$ fully analytically. Specifically, the elements of the scattering matrix $S$ for this setup read as follows
\begin{align}
	S_{11} & =S_{22} = -\frac{e^{-2 \mathrm{i} k L} \left(-1+e^{4 \mathrm{i} k L n}\right) \left(n^2-1\right)}{e^{4 \mathrm{i} k L n} (n-1)^2-(n+1)^2},\\
	S_{12} & =S_{21} =-\frac{4 e^{2 \mathrm{i} k L (n-1)} n}{e^{4 \mathrm{i} k L n} (n-1)^2-(n+1)^2},
\end{align}
where $k$ is the wavenumber of the incident field. Note that in contrast to the main text all formulas here are written in terms of the refractive index $n=\sqrt{\varepsilon}$ instead of the dielectric function $\varepsilon$ in order to simplify the notation.
\subsection{$Q_n$ for Refractive Index Variation (in 1D)}
To better understand $Q_\varepsilon$, the GWS-operator associated to the total intensity inside a target, we first consider the GWS-operator we get when taking the derivative of the scattering matrix with respect to the refractive index $n$ of the barrier. At the end of this section we also show the results for $\alpha=\varepsilon$. The components of $Q_n=-\mathrm{i}S^{-1}\mathrm{d}S/\mathrm{d}n$ are
\begin{align}
Q_{n,11} & = Q_{n,22} = \frac{-8 k L \left(n^3+n\right)-2 \left(n^2-1\right) \sin (4 k L n)}{-\left(n^2+6\right) n^2+\left(n^2-1\right)^2 \cos (4 k L n)-1},\\
Q_{n,12} & = Q_{n,21} = \frac{-8 k L n \left(n^2-1\right) \cos (2 k L n)-4 \left(n^2+1\right) \sin (2 k L n)}{-\left(n^2+6\right) n^2+\left(n^2-1\right)^2 \cos (4 k L n)-1}.
\end{align}
The corresponding eigenvalues $\theta_n^{1,2}$ and eigenstates $\vec{u}_n^{1,2}$ then follow as
\begin{align}
	\theta^1_n &= \frac{-4 k L n-2\sin (2 k L n)}{\left(n^2-1\right) \cos (2 k L n)-n^2-1}, &
	\vec{u}^1_n & =  \frac{1}{\sqrt{2}}\left(
	\begin{array}{cc}
	1 \\ 1
	\end{array}
	\right),\\
	\theta^2_n &= \frac{4 k L n-2\sin (2 k L n)}{\left(n^2-1\right) \cos (2 k L n)+n^2+1}, &
	\vec{u}^2_n & =  \frac{1}{\sqrt{2}}\left(
	\begin{array}{cc}
	-1 \\ 1 
	\end{array}
	\right).
\end{align}
The integrated intensities of the scattering states inside the barrier read as
\begin{align}
	I_1 & \equiv \int_{-L}^{L} \absq{\psi\left( \vec{u}^1_n\right)} \mathrm{d}x = \frac{-4 k L n-2\sin (2 k L n)}{k n \left[\left( n^2-1\right) \cos (2 k L n)-n^2 -1\right]}  = \frac{\theta^1_n}{kn},\\
	I_2 & \equiv \int_{-L}^{L} \absq{\psi\left( \vec{u}^2_n\right)} \mathrm{d}x = \frac{4 k L n-2 \sin (2 k L n)}{k n \left[\left(n^2-1\right) \cos (2 k L n)+n^2+1\right]} = \frac{\theta^2_n}{kn}.
\end{align}
This result shows that there is a strict linear relation between the stored intensity in the designated scattering region and the eigenvalues of $Q_n$ which enables us to tune the intensity inside the scatterer based on the choice of $Q_n$-eigenstates (or a superposition of such eigenstates). Taking eigenstates with large eigenvalues results in scattering states that focus into the target region.\\
To calculate $Q_\varepsilon$ we make use of  $\mathrm{d}/\mathrm{d}n = \mathrm{d}/\mathrm{d}\varepsilon \cdot \mathrm{d}\varepsilon/\mathrm{d}n = 2 n \mathrm{d}/\mathrm{d}\varepsilon $, which directly leads to the following expression
\begin{equation}
\frac{I_1}{\theta^1_\varepsilon} =\frac{I_2}{\theta^2_\varepsilon} = \frac{2}{k}.
\end{equation}
\subsection{$Q_x$ for Target Displacement (in 1D)}
To construct $Q_x$ we take the derivative of the scattering matrix with respect to the position of the barrier. To obtain the eigenvalues associated to the momentum-transfer onto the target we make use of $Q_x = k_\mathrm{in} - S^\dagger k_\mathrm{out} S$ \cite{ambichl_focusing_2017}. This formula comes from the fact that for only one scatterer (or if we move all scatterers) $Q_x$ measures the momentum difference between incoming and outgoing waves. The matrices $k_\mathrm{in}$ and $k_\mathrm{out}$ read as follows
\begin{equation}
	k_{\mathrm{in}} = \left(
	\begin{array}{cc}
	k & 0 \\ 0 &-k
	\end{array}
	\right) = -k_{\mathrm{out}}.
\end{equation}
The components of $Q_x$ then read
\begin{align}
	-Q_{x,11} &=  Q_{x,22} = \frac{1}{\Gamma}\left[4k\left(n^2-1\right)^2\sin^2\left(2kLn\right)\right],\\
	-Q_{x,12} & = Q_{x,21} = \frac{1}{\Gamma}\left[8\mathrm{i}kn\left(n^2-1\right)\sin\left(2kLn\right)\right],
\end{align}
with $\Gamma \equiv -\left(n^2-1\right)^2\cos\left(4kLn\right)+n^2\left(n^2+6\right)+1 >0 \enskip\forall n,k,L$. The solution of the eigenproblem associated with $Q_x$ is
\begin{align}
	\theta^{1,2}_x & = \pm \frac{\sqrt{8}k\left(n^2-1\right)\sin\left(2kLn\right)}{\sqrt{\Gamma}} ,\\
	\vec{u}^{1,2}_x &=\mp \frac{1}{N_{1,2}}\left(\frac{-\mathrm{i}}{4n}\left[\sqrt{2\Gamma} \pm 2\left(n^2-1\right)\sin\left(2kLn\right)\right], 1\right)^T,
\end{align}
where $N_{1,2} = |\vec{u}^{1,2}_x|$ is the norm of the eigenvectors. We then calculate the quantity $F_{1,2}$, which we will show to be proportional to the momentum transfer onto the target:
\begin{align}
	F_{1,2} & \equiv  \absq{\psi\left( \vec{u}^{1,2}_x, x= - L\right)}   -  \absq{\psi\left( \vec{u}^{1,2}_x, x= +L\right)}  \nonumber\\
	&= \mp\frac{4\sqrt{2}\sin\left(2kLn \right)}{\sqrt{\Gamma}},
\end{align}
where the first argument in the parentheses tells us which eigenvector of $Q_x$ is used in the calculation. We then compare these two quantities with the eigenvalues $\theta_x^{1,2}$ of $Q_x$ and see that
\begin{equation}\label{eq:vers}
	\frac{\theta_x^{1}}{F_1} = \frac{\theta_x^{2}}{F_2} = -\frac{k\left(n^2-1\right)}{2}.
\end{equation}
This result tells us that there is a strict linear relation between the eigenvalues of $Q_x$ and the difference in intensities at the left and right boundary of the barrier, $F_i$. The eigenvalues $\theta^{1,2}_x$ are equal to the momentum difference between incoming and outgoing waves, i.e., $\Delta k =\theta_x$ \cite{ambichl_focusing_2017}. Conservation of momentum then tells us that a momentum of $\Delta k$ is transferred to the target. We work in the stationary case, thus this momentum transfer onto the target is the same at all times. The average force, $F$, experienced by the target is the momentum transfer per unit time interval and since the momentum transfer is stationary, it is equal to the force. This implies that the quantities $F_i$ we have defined above are proportional to the force the target experiences from the scattering of the wave. This enables us to tune the degree of force exerted onto the target and even control the direction of it.
\subsection{$Q_R$ for Radial Change (in 1D)}
The one-dimensional analogue to the radius $R$ of a circle is the length $L$ of a region with raised refractive index. We therefore investigate the behaviour of $Q_L = -\im S^{-1}\mathrm{d}S/\mathrm{d}L$, whose elements are given by
\begin{align}
	Q_{L,11} & = Q_{L,22} = -\frac{2 k \left(n^2-1\right) \left[3 n^2+\left(n^2-1\right) \cos (4 k L n)+1\right]}{-\left(n^2+6\right) n^2+\left(n^2-1\right)^2 \cos (4 k L n)-1},\\
	Q_{L,12} & = Q_{L,21} = -\frac{8 k n^2 \left(n^2-1\right) \cos (2 k L n)}{-\left(n^2+6\right) n^2+\left(n^2-1\right)^2 \cos (4 k L n)-1}.
\end{align}
The solution of the corresponding eigenproblem yields
\begin{align}
	\theta^1_L & = \frac{-4 k \left(n^2-1\right) \cos ^2(k L n)}{\left(n^2-1\right) \cos (2 k L n)-n^2-1}, &
	\vec{u}^1_L & = \frac{1}{\sqrt{2}}\left(
	\begin{array}{cc}
	1 \\ 1
	\end{array}
	\right),\\
	\theta^2_L & = \frac{4 k \left(n^2-1\right) \sin ^2(k L n)}{\left(n^2-1\right) \cos (2 k L n)+n^2+1}, &
	\vec{u}^2_L & = \frac{1}{\sqrt{2}}\left(
	\begin{array}{cc}
	-1 \\ 1
	\end{array}
	\right).
\end{align}
The intensities at the boundaries of the barrier are
\begin{align}\label{eq:QR}
	P_1 & \equiv  2\absq{\psi\left( \vec{u}^1_L, x= \pm L\right)} = \frac{-8 \cos ^2(k L n)}{\left(n^2-1\right) \cos (2 k L n)-n^2-1} = \frac{2\theta_L^1}{k\left(n^2-1\right)},\\
	P_2 & \equiv 2\absq{\psi\left( \vec{u}^2_L, x= \pm L\right)} = \frac{8 \sin ^2(k L n)}{\left(n^2-1\right) \cos (2 k L n)+n^2+1} = \frac{2\theta_L^2}{k\left(n^2-1\right)},
\end{align}
for each eigenstate of $Q_L$.
In contrast to $Q_n$ whose eigenvalues represent the integrated intensity over the target region, the above result shows that the eigenvalues of $Q_L$ correspond to the intensities at the target's boundary which can be again tuned by choosing a certain $Q_L$-eigenstate (or a superposition of such eigenstates).

\section{Derivation of Eq. 2}
In the following we will sketch the derivation of Eq. 2 from the main text involving the expectation value of the generalised Wigner-Smith (GWS) operator with respect to some arbitrary input vector $\ket{\chi}$. All conventions are taken from \cite{PADiss} (see also \cite{fyodorov_statistics_1997, rotter_light_2017}). The scattering matrix $S$ for a system described by the Helmholtz equation can be written as 
\begin{equation}
	S = -\mathbbm{1} + 2\im V^\dagger G V, \enskip \mathrm{where} \enskip G=\left(\Delta + U(\vec{x}) +\im VV^\dagger \right)^{-1}\enskip \mathrm{and} \enskip U(\vec{x}) = k^2 \varepsilon(\vec{x}),
\end{equation}
where $G$ is the system's Green's function, $\Delta$ is the Laplacian, $U$ determines the scattering environment and $V$ describes the coupling to the leads. The derivative of the scattering matrix can then be written as,
\begin{align}
	\de{S}{\alpha} & = -2 \im V^\dagger G \de{U}{\alpha}GV,
\end{align}
where we have assumed that only the scattering environment $U$ depends on $\alpha$, but not the coupling matrix $V$. Under the further assumption of a unitary $S$-matrix the following useful identity was derived in \cite{PADiss} for the Green's function $G$,
\begin{equation}
	-2\im G^\dagger VV^\dagger G = G - G^\dagger.
\end{equation}
With a unitary scattering matrix we can then write $Q_\alpha = -\im S^{-1} \de{S}{\alpha}$ as,
\begin{align}
	Q_\alpha & = - 2 S^\dagger V^\dagger G \de{U}{\alpha} G V  \\
	& = 2\left(\mathbbm{1} + 2 \im V^\dagger G^\dagger V\right)V^\dagger G \de{U}{\alpha} G V = 2 V^\dagger \left( G-G+G^\dagger\right) \de{U}{\alpha} G V\\
	Q_\alpha& = 2 V^\dagger G^\dagger \de{U}{\alpha} G V.
\end{align}
The wavefunction inside the scattering system for an arbitrary input vector $\ket{\chi}$ can be written as \cite{PADiss},
\begin{equation}
	\ket{\psi_\chi} = 2 \im G V \ket{\chi},
\end{equation}
resulting in the following expression for the expectation value of $Q_\alpha$,
\begin{equation}
	\bra{\chi}Q_\alpha \ket{\chi} = \frac{1}{2} \bra{\psi_\chi}\de{U}{\alpha} \ket{\psi_\chi}.
\end{equation}
This novel result now allows us to extract information about the local wave intensity inside the scattering medium out of the asymptotic information encoded in the scattering matrix $S$.\\\indent
As a simple test of the above relation, we take the derivative of $U$ with respect to $k$ and we get
\begin{equation}
	Q_k = 4k V^\dagger G^\dagger \varepsilon G V \equiv Q_d, 
\end{equation}
which is exactly the result for the dwell-time operator as given in \cite{PADiss, rotter_light_2017}. Keeping in mind that the dwell time operator and the Wigner-Smith time-delay operator are the same for a vanishing $k$-dependence of $V$ \cite{PADiss, rotter_light_2017}, we have thus shown that our new relation is perfectly consistent with this established case.

\section{Linear Relations in Two Dimensions}
\begin{figure}[t!]
	\centering
	\includegraphics[width=\textwidth]{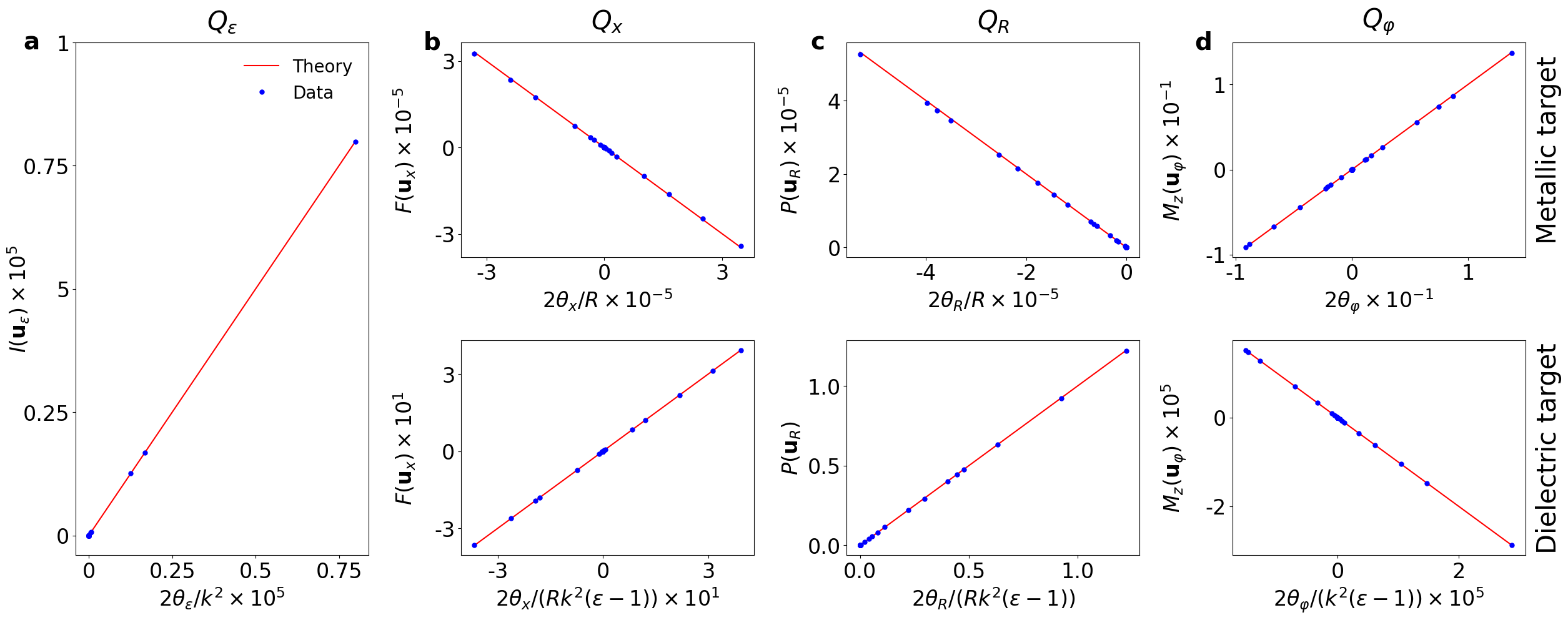}
	\caption{Linear relations of the GWS-eigenvalues with the quantities defined in Eqs.~(\ref{eq:dn})-(\ref{eq:tor}), for $N=20$ propagating modes and a wavelength of $\lambda\approx0.1W$, where $W$ is the width of the waveguide. The simulation is carried out in the waveguide geometry depicted in Fig.~1 of the main text. The simulated data is depicted by the blue dots and shows an excellent agreement with the theoretical predictions (red line) for all cases (without any free parameters). \textbf{a}, Linear relation between $\theta_\varepsilon$ and the stored intensity $I$ for a Teflon ($\varepsilon=2.0736$) target (radius $R=0.0165W$).
	\textbf{b}, Upper (lower) plot shows the linear relation between $\theta_x$ and $F$ for a metallic (Teflon) target ($R=0.0825W$).
	\textbf{c}, Upper (lower) plot shows the linear relation between $\theta_R$ and $P$ for a metallic (Teflon) target.
	\textbf{d}, Upper (lower) plot shows the linear relation between $\theta_\varphi$ and $M_z$ for a metallic (Teflon) square target (side length~$0.165W$).
	}
	\label{fig:linear_realtions}
\end{figure}
Building up on the insights gained in the previous section, we can now prove the linear relations written down in Eqs.~(3) and (4) of the main text. We start by proving the relation between $\theta_\varepsilon$ and the intensity integrated over the target's area. Taking the derivative of the scattering environment with respect to the dielectric constant of a single scatterer results in $\mathrm{d}U/\mathrm{d}\varepsilon = k^2 A_{\mathrm{scat}}\left(\vec{r}\right)$, where $A_{\mathrm{scat}}$ parameterises the area of the scatterer and is equal to one inside the target and zero outside. This then gives
\begin{equation}
	I \left(\ket{u_\varepsilon^i}\right) \equiv \int \lvert \psi\left(\ket{u_\varepsilon^i}\right) \rvert^2 \mathrm{d} A = \frac{2\theta_\varepsilon^i}{k^2},
	\label{eq:dn}
\end{equation}
where the integral is performed over the area $A$ of the target scatterer. This equation is numerically confirmed in Fig.~\ref{fig:linear_realtions}a and equips us with a tool to control the wave intensity inside a target's area -- ranging from no intensity to the theoretical maximum. \\\indent
Next, we derive in detail the linear relation between the eigenvalues of $Q_R$ and the pressure, $P\left(\ket{u_R^i} \right) \equiv \int_{0}^{2\pi} \absq{\psi\left(\rho=R\right)} \mathrm{d} \varphi$, applied to the target. We start by considering a dielectric target, for which $\mathrm{d}U/\mathrm{d}R = k^2(\varepsilon-1) \delta\left(\rho-R\right)$. This leads us to
\begin{align}
	\frac{1}{2}\braket{\psi | \de{U}{R} | \psi} & = \frac{k^2\left(\varepsilon-1\right)}{2}\int_{0}^{2\pi}\int_{0}^{\infty} \absq{\psi\left(\rho,\varphi\right)}\rho\,  \delta\left(\rho-R\right) \mathrm{d}\rho\mathrm{d}\varphi \\
	& = \frac{k^2\left(\varepsilon-1\right)R}{2} \int_{0}^{2\pi} \absq{\psi\left(\rho=R,\varphi\right)} \mathrm{d}\varphi\\
	& = \frac{k^2\left(\varepsilon-1\right)R}{2} P\left(\ket{u_R^i}\right).
\end{align}
This derivation provides us with the final result that the pressure is in a linear relation to the eigenvalues of $Q_R$
\begin{equation}
	P\left(\ket{u_R^i} \right) \equiv \int_{0}^{2\pi} \absq{\psi\left(\rho=R\right)} \mathrm{d} \varphi = \frac{2 \theta_R^i}{Rk^2(\varepsilon-1)}.
\end{equation}
We confirm this equation numerically in Fig.~\ref{fig:linear_realtions}c. To prove a linear relation between the eigenvalues of $Q_R$ and the transferred pressure in the case of a metallic target with perfect conductance we first consider the simple example of an infinitely long metallic rod of circular cross-section in vacuum. Effectively this configuration can be reduced to a two-dimensional problem.
Due to the simplicity of the geometry, different incoming cylindrical wave modes do not mix, therefore the scattering matrix $S$, which in this case is just a unitary diagonal reflection matrix $r$, reads
\begin{equation}
	\left[r\right]_{nn} = -\frac{\mathrm{H}_n^{(2)}\left(kR\right)}{\mathrm{H}_n^{(1)}\left(kR\right)},
\end{equation}
where $\mathrm{H}_n^{(1)}$ and $\mathrm{H}_n^{(2)}$ are the Hankel functions of the first and second kind, respectively. The knowledge of the reflection matrix then allows us to compute the GWS-operator $Q_R$, whose elements on the diagonal read
\begin{equation}
	\left[Q_R\right]_{nn} = -\frac{4}{\pi R\left[\mathrm{J}_n\left(kR\right)^2 + \mathrm{Y}_n\left(kR\right)^2\right]},
\end{equation}
where $J_n$ and $Y_n$ are the Bessel functions of the first and second kind, respectively. The wavefunction around the metallic rod comprised of the incident and scattered wave is
\begin{equation}
	\psi\left(\rho,\varphi\right) = \frac{1}{2}\sum_{n=-\infty}^{\infty} b_n e^{\im n \varphi}\left[\mathrm{H}_n^{(2)}\left(k\rho\right) + \left[r\right]_{nn} \mathrm{H}_n^{(1)}\left(k\rho\right)\right],
\end{equation}
where $b_n$ are the components of the input vector. This leads to the following expression in the case when $b_n=1$ for only one $n$ and zero for all other
\begin{equation}
	P \equiv \int_{0}^{2\pi} \absq{\partial_\rho \psi \left(\rho=R\right)} \mathrm{d}\varphi = \frac{8}{\pi R^2 \left[\mathrm{J}_n\left(kR\right)^2 + \mathrm{Y}_n\left(kR\right)^2\right]}.
\end{equation}
We thus find
\begin{equation}
	P\left(\ket{u_R^i}\right) = -\frac{2\theta_R^i}{R},
\end{equation}
which we confirm numerically in Fig.~\ref{fig:linear_realtions}c for a circular metallic target inside a waveguide featuring the same disorder as in Fig.~1 of the main manuscript. The reason why our derivation of the proportionality constant also works for an arbitrary environment as wells as an arbitrary cross-section, is that the proof of the GWS-operator tells us that its eigenvalues $\theta_\alpha$ are in a linear relation with the local wave intensity around the target, irrespectively of the surrounding scattering environment and whether these are waveguide walls or a disordered medium.\\\indent
Next we consider the eigenvalues of $Q_x$, the GWS-operator we get when the parameter $\alpha$ considered is the longitudinal position $x$ of a target. This problem was originally considered in \cite{ambichl_focusing_2017}, although a proof was only given for the case of longitudinally moving the entire scattering system. Here we close this gap and prove a linear relation between the eigenvalues and the momentum transferred also for a single target inside a disordered medium, using the same strategy as in the preceding paragraph. For a dielectric circular target with radius $R$, the result reads
\begin{equation}
	F\left(\ket{u^i_{\hat{\vec{n}}}}\right)\equiv \hat{\vec{n}}\cdot \int_{0}^{2\pi} \left(\begin{array}{cc} \cos\varphi \\ \sin\varphi \end{array}\right) \absq{\psi\left(\rho=R\right)} \mathrm{d}\varphi = \frac{2\theta^{i}_{\hat{\vec{n}}}}{Rk^2(\varepsilon-1)},
\end{equation}
where the direction of the shift and the corresponding momentum transfer is generalised to an arbitrary direction, which is parametrised by the unit vector $\hat{\vec{n}}$ and the integral is performed along the boundary of the circular target (a generalization to arbitrary target shapes is also possible). In the presence of a metallic target the result is
\begin{equation}
	F\left(\ket{u_{\hat{\vec{n}}}^i}\right) \equiv \hat{\vec{n}}\cdot  \int_{0}^{2\pi}  \left(\begin{array}{cc} \cos\varphi \\ \sin\varphi \end{array}\right) \absq{\partial_\rho\psi\left(\rho=R\right)} \mathrm{d}\varphi = -\frac{2\theta_{\hat{\vec{n}}}^i}{R}.	
\end{equation}
Numerical results confirming these relations for $\hat{\vec{n}}=(1,0)^T$ can be found in Fig.~\ref{fig:linear_realtions}b.\\\indent
The same approach also works when considering the eigenvalues of $Q_\varphi$ that are in a linear relation with the torque transferred to the target. In the case of a dielectric target this relation reads
\begin{align}
	M_z\left(\ket{u_\varphi^i}\right) \equiv \int_\mathcal{C} \left[\vec{m}_\perp \left(\vec{c}\right) \times \vec{n}\left(\vec{c} \right) \absq{\psi\left(\vec{c}\right)}\right]_z \mathrm{d}s = -\frac{2\theta^i_\varphi}{k^2(\varepsilon-1)},
\label{eq:tor_d}
\end{align}
where the integral is taken along the target's boundary described by the curve $\mathcal{C}$ and parametrised by $\vec{c}$. The expression $\vec{n}\left(\vec{c} \right) \absq{\psi\left(\vec{c}\right)}$ denotes the normal force at every point of the boundary excised by the electric field and $\vec{m}_\perp\left(\vec{c}\right)$ is the part of the distance from the boundary to the target's center of mass that is normal to $\vec{n}$, i.e., the lever. We find that in the presence of a metallic target, the eigenvalues $Q_\varphi$ are proportional to
\begin{align}
	M_z\left(\ket{u_\varphi^i}\right) \equiv \int_\mathcal{C} \left[\vec{m}_\perp \left(\vec{c}\right) \times \vec{n}\left(\vec{c} \right) \absq{\partial_{\vec{n}} \psi\left(\vec{c}\right)}\right]_z \mathrm{d}s = 2\theta^i_\varphi,
	\label{eq:tor}
\end{align}
where $\vec{n}\left(\vec{c} \right) \absq{\partial_{\vec{n}} \psi\left(\vec{c}\right)}$ denotes the normal force at every point of the boundary excised by the electric field. These two equations are numerically verified in Fig.~\ref{fig:linear_realtions}d for the case of a square target.

\section{Applying Pressure with the GWS-Operator}

\begin{figure}[t!]
	\centering
	\includegraphics[width=\textwidth]{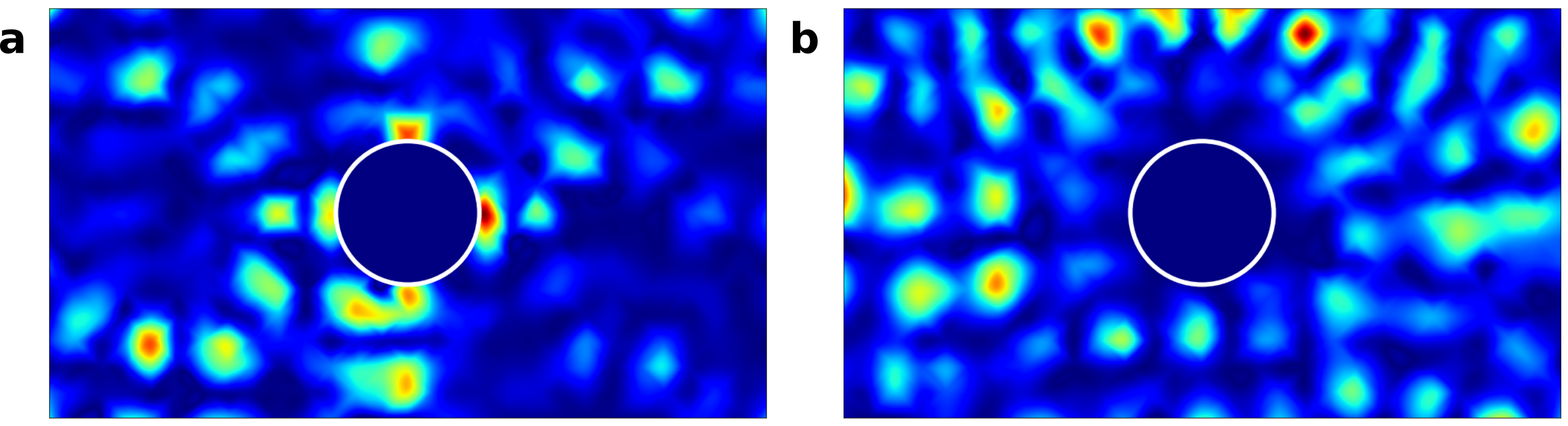}
	\caption{\textbf{a}, Measured spatial intensity distribution for the scattering state created by the eigenvector $\ket{u_R^\mathrm{max}}$ of $Q_R$ corresponding to the largest eigenvalue $|\theta_R^\mathrm{max}|\approx 709$. A strong focus on the metallic target's boundary (radius $R=14$ mm) can clearly be observed in the experimental data, verifying our prediction that these states control the applied pressure. \textbf{b}, To contrast this we show here the experimentally obtained intensity distribution for a state corresponding to a small eigenvalue $|\theta_R^\mathrm{small}| \approx 86$.}
	\label{fig:press}
\end{figure}

In this section we look at the GWS-operator $Q_R=-\im S^{-1} \mathrm{d}S/\mathrm{d}R$, where $R$ is the radius of a metallic circular scatterer (our approach also works for more complicated target shapes). For the corresponding eigenvectors $\ket{u_R^i}$ we analytically showed in the preceding section that there exists a linear relation between the radiation pressure applied to such a metallic target and the corresponding eigenvalue $\theta_R^i$ of the following form:
\begin{equation}
	P\left(\ket{u_R^i}\right) \equiv \int_{0}^{2\pi} \absq{\partial_\rho\psi\left(\rho=R\right)} \mathrm{d}\varphi = -2\theta_R^i/R,
\end{equation}
where the integral is along the boundary of the circular scatterer with radius $R$ and $\psi$ is the electric field distribution of the corresponding eigenstate $\ket{u_R^i}$. In other words, the eigenstates $\ket{u_R^i}$ of $Q_R$ are characterised by a well-defined radiation pressure $P$ that they apply to the dielectric target scatterer. The pressure applied to a chosen target can thus be maximised by injecting eigenstates $\ket{u_R^\mathrm{max}}$ corresponding to the largest eigenvalues $\theta_R^\mathrm{max}$. The experimentally measured scattering state of the eigenstate associated to the largest eigenvalue can be seen in Fig.~\ref{fig:press}a, where the experiment was carried out in a waveguide featuring a disorder (see Fig. 1). We can clearly see that the state has a strong intensity build-up on all sides of the boundary of the chosen scatterer. Constructing $Q_R$ with a dielectric scatterer (rather than a metallic one) as a target yields the following linear relation between the radiation pressure $P$ and the eigenvalues $\theta_R^i$,
\begin{equation}
	P\left(\ket{u_R^i}\right) \equiv \int_{0}^{2\pi} \absq{\psi\left(\rho=R\right)} \mathrm{d}\varphi = \frac{2\theta_R^i}{Rk^2\left(\varepsilon-1\right)}.
\end{equation}

\section{Comparing the GWS-Method and the Field Matrix Method}
\begin{figure}[t!]
	\centering
	\includegraphics[width=\textwidth]{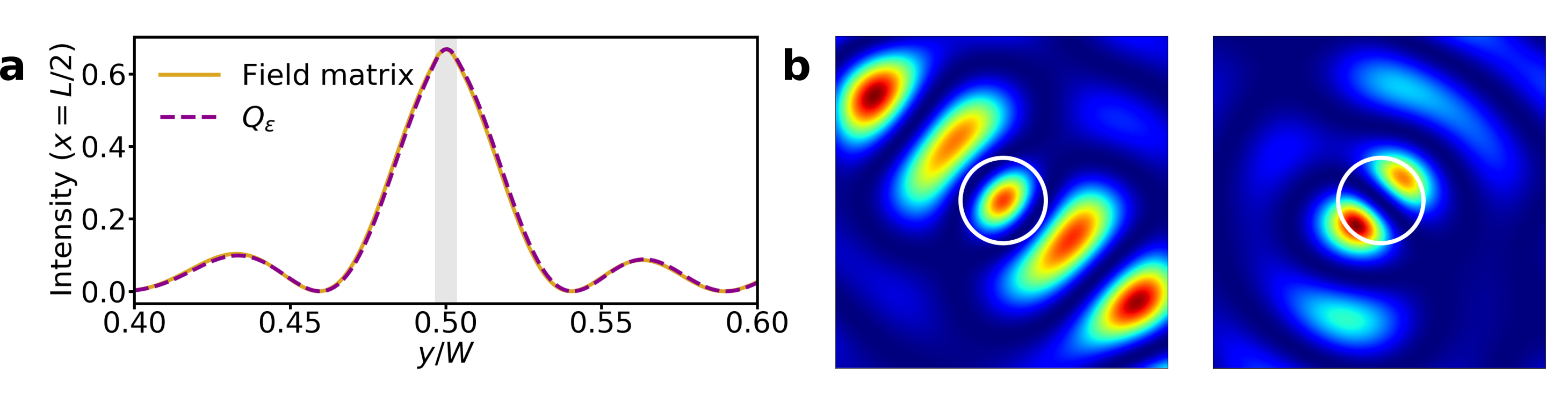}
	\caption{\textbf{a}, Simulated intensity of the state calculated with the field matrix method \cite{Cheng:14}, which produces an optimal focus in a single point, and the highest $Q_\varepsilon$-eigenstate at the center of the target ($x=L/2$) as a function of the transverse coordinate $y$ (the grey-shaded region marks the extension of the target scatterer). In these simulations we use the same geometry as in the experiment, featuring a disorder, but with a frequency of 30 GHz resulting in 20 propagating modes and a wavelength of 1 cm. The diameter of the target is 0.69 mm which corresponds to 10\% of the wavelength inside the target. The resulting state is identical to the highest $Q_\varepsilon$-eigenstate, confirming our prediction that we are able to achieve optimal focus. \textbf{b}, Comparison of the field matrix result (left) with a focus point in the middle of the target to the highest $Q_\varepsilon$-eigenstate (right) when $\varepsilon$ is varied inside the entire target with radius 10.4 mm. The two peaks produced by the GWS-method results in a two-fold increase of the integrated intensity in the target as compared to the one peak produced by the field matrix method.}
	\label{fig:fmm}
\end{figure}

Here we verify explicitly that our micro-manipulation protocol is indeed optimal by comparing it to a procedure proposed in \cite{Cheng:14}. There the authors achieve a focus onto a point inside a scattering medium by utilizing the field matrix $e(x)$, which relates the field at a depth $x$ inside the scattering medium to the incident field in the waveguide's leads. The components of the field matrix $e_{ab}(x)$ connect the field in channel $b$ at depth $x$ with the field in channel $a$ at the waveguide's leads, i.e., $E_b(x) = e_{ba}(x) E_a$, where $E_a$ and $E_b(x)$ are the fields in channels $a$ and $b$. Optimal focus, i.e., maximizing $\lvert E_\beta(x)\rvert$, at a target point $\beta$ can now be achieved by shaping the incident wavefront as
\begin{equation}
	E_a^{\mathrm{opt}} = e^*_{\beta a}(x)I_\beta^{-1/2}(x),
\end{equation}
with $I_\beta (x) = \sum_a \lvert e_{\beta a}(x)\rvert^2$ (for a proof see \cite{PhysRevLett.101.120601}). In Fig.~\ref{fig:fmm}a we compare the focus achieved by the field matrix method with the one achieved with the highest $Q_\varepsilon$-eigenstate for a circular target scatterer with a very small diameter $D$ (a very small target allows us to converge to the limit of focusing on a point, as considered in the field-matrix method). Our comparison, indeed, shows that the highest $Q_\varepsilon$-eigenstate is indistinguishable from the one constructed by the field matrix method, i.e., they both deliver the optimal focus in form of a single peak in the center of the scatterer. In contrast to the field matrix method which, however, can only focus onto a single point, the GWS-concept also enables optimal focusing into an extended area of arbitrary size. Focusing into an extended area requires more than one peak such that the field matrix method would then already have to know where exactly all these peaks have to lie, which is equivalent to already knowing the state that leads to optimal focus in the first place. To demonstrate that the GWS-concept is able to focus in an extended area we show in Fig.~\ref{fig:fmm}b the spatial intensity distribution of the highest $Q_\varepsilon$-eigenstate around a target scatterer with a larger area. In order to maximise the stored intensity, $Q_\varepsilon$ constructs the highest eigenstate such that two peaks fit inside the scatterer, resulting in a two-fold increase of the stored intensity as compared to the field-matrix solution with just one peak in the center of the target.

\section{Incomplete Access to Scattering Channels}
\begin{figure}[!t]
	\centering
	\includegraphics[width=\textwidth]{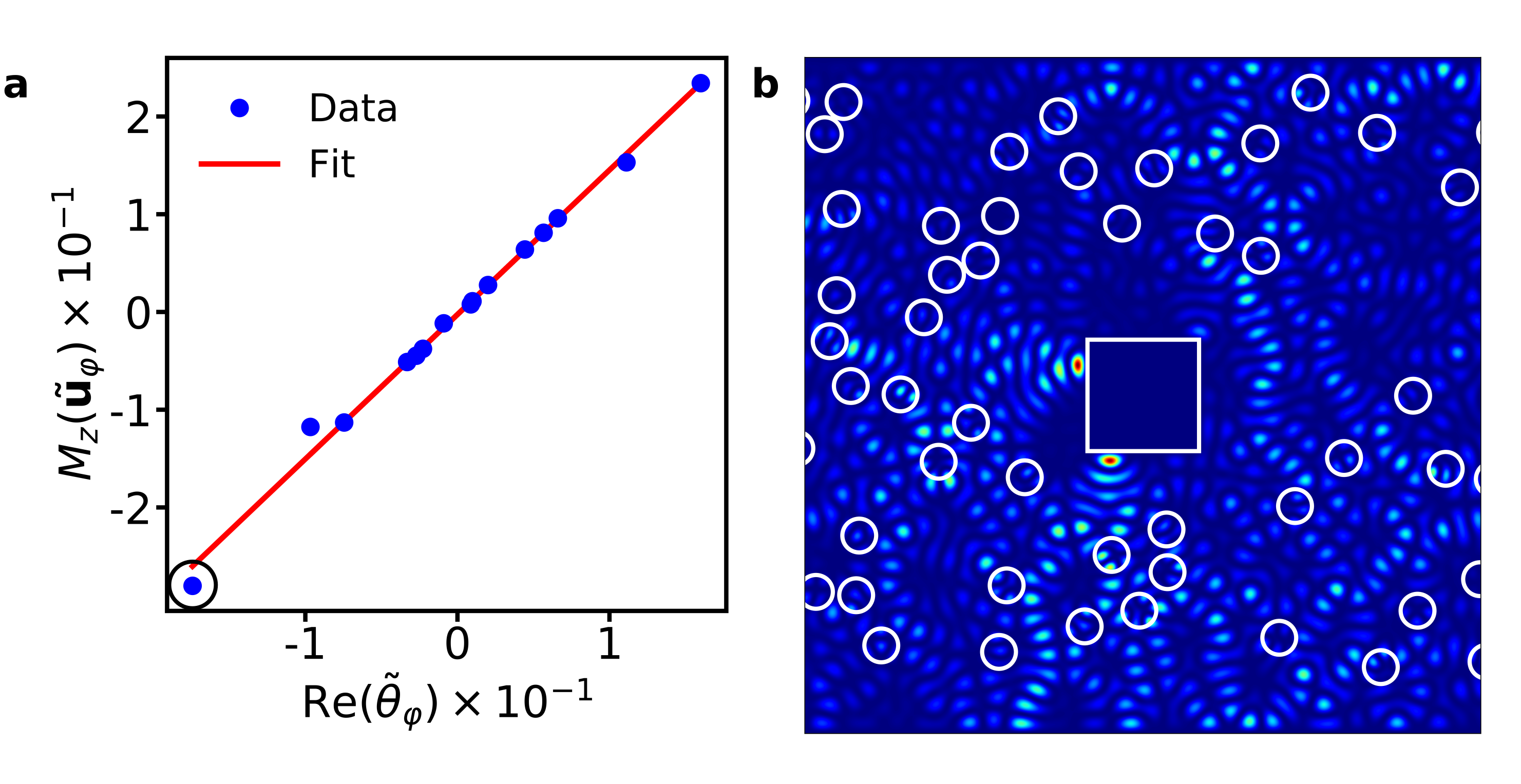}
	\caption{\textbf{a}, Linear relation between the torque $M_z\left(\ket{\tilde{u}_\varphi}\right)$ and the real part of $\tilde{q}_\varphi$~-~eigenvalues $\mathrm{Re} \left(\tilde{\theta}_\varphi\right)$ in the case of incomplete access to scattering channels. The target here is a metallic square embedded into a disorder of 300 randomly distributed Teflon scatterers, where the whole system has a reflectivity of $R \approx 0.87$. Despite the limitations an approximately linear behaviour can be clearly seen. Here, $N=40$ propagating modes were used, where the $M=10$ highest ones are neglected in the calculation of $\tilde{q}_\varphi$ to simulate a low numerical aperture. In the SVD-procedure, we project onto the 15 highest reflecting states, which results in a correlation coefficient of $r=0.997$. The black circle marks the eigenvalue with the largest absolute value of $\mathrm{Re} \left(\tilde{\theta}_\varphi\right)$ whose corresponding eigenstate is shown in b. \textbf{b}, Spatial intensity distribution of the $\tilde{q}_\varphi$-eigenstate corresponding to the largest absolute value of $\mathrm{Re} \left(\tilde{\theta}_\varphi\right)$ which focuses onto the edges of the square target in order to maximise the transferred torque.}
	\label{fig:nonidealistic_qphi}
\end{figure}
In the main document we explore different realisations of GWS-operators for the ideal case where we have access to the full scattering matrix, $S$. Since it is experimentally challenging to measure the entire scattering matrix, we numerically show that our approach also works in the regime of a sub-unitary $S$-matrix. For this case we consider the same waveguide ($L=6W$) as in the main document but fill it with $300$ randomly distributed Teflon scatterers  ($R=0.025W$, $n=1.44$) and a square metallic target (side length $0.165W$) at $x$-position $L/6$. Instead of the full scattering matrix, we now take only the reflection matrix $r$ for the calculation of the GWS-operator, since in the experiment both transmission and reflection measurements are not always possible. Experiments also suffer from a low numerical aperture (LNA), which we take into account in our simulation by removing the $M$ highest modes from $r$ such that we are left with a $(N-M)\times (N-M)$ matrix $r_\mathrm{LNA}$, where $N$ is the number of modes. When we use only the reflection matrix to construct an operator $q_\varphi=-\mathrm{i} r^{-1}\mathrm{d}r/\mathrm{d}\varphi$ its eigenvalues turn out to be $\theta_\varphi^i = \left(M_z(\vec{u}_\varphi^i) + \mathrm{i} \vec{u}_\varphi^{i,\dagger}t^\dagger \mathrm{d}t/\mathrm{d}\varphi \vec{u}_\varphi^i\right)/(2\langle r^\dagger r\rangle)$ \cite{ambichl_focusing_2017}, where $\langle r^\dagger r\rangle$ is the global reflectance of the eigenstate and the loss of information manifests itself in complex eigenvalues caused by the second term containing the transmission matrix $t$. From this expression we may thus conclude that our procedure based on the reflection matrix only will work better for higher total reflection and for the case that the target is placed at the beginning of the waveguide (placing it on the end would result in far less intensity reaching it). In order to still maintain a correlation between the eigenvalues of $q_\varphi$ and the total torque acting on the target scatterer, we also have to restrict ourselves to channels that are strongly reflecting to minimise the term with the transmission matrix $t$. We achieve this by using a singular-value decomposition of $r_\mathrm{LNA} = U\Sigma V^\dagger$, where the matrices $U$ and $V$ contain the column wise the left and right singular vectors (not to be confused with the scattering environment and the coupling matrix, respectively) and the matrix $\Sigma = \mathrm{diag}\left(\left\{ \sigma_n \right\}\right)$ contains the singular values on its diagonal. In order to project on the highly reflecting channels we pick a certain subset of large singular values $\tilde{\Sigma} = \mathrm{diag}\left(\left\{ \tilde{\sigma}_n \right\}\right)$ as well as the associated left and right singular vectors $\tilde{U}$ and $\tilde{V}$. Equipped with these matrices we can construct an effective inverse $r_\mathrm{LNA}^{-1} = \tilde{V}\left(\tilde{U}^\dagger r_\mathrm{LNA} \tilde{V}\right)^{-1} \tilde{U}^\dagger$ and also project the derivative of $r_\mathrm{LNA}$ onto this subspace with the proper projection operators $P_{\tilde{U}} = \tilde{U} \tilde{U}^\dagger$ and $P_{\tilde{V}} = \tilde{V} \tilde{V}^\dagger$ \cite{ambichl_focusing_2017}. As a final result we obtain:
\begin{equation}
	\tilde{q}_\varphi = -\mathrm{i} \tilde{V}(\tilde{U}^\dagger r_\mathrm{LNA}\tilde{V})^{-1} \tilde{U}^\dagger\tilde{U}\tilde{U}^\dagger\de{r_\mathrm{LNA}}{\varphi} \tilde{V} \tilde{V}^\dagger.
\end{equation}
In Fig.~\ref{fig:nonidealistic_qphi}a one can see that Eq.~\eqref{eq:tor} is almost perfectly fulfilled, i.e., our protocol works even for the case where only parts of the scattering matrix are available. This, however, comes with the caveat that the system needs to be highly reflecting (transmitting) when we work only with the reflection (transmission) matrix. Fig.~\ref{fig:nonidealistic_qphi}b shows the spatial intensity distribution of the $\tilde{q}_\varphi$-eigenstate which transfers the maximal amount of torque.

\bibliographystyle{ieeetr}
\bibliography{bib1,bib2,bib3,bib4,bib5}

\begin{thebibliography}{1}

\bibitem{ambichl_focusing_2017}
P.~Ambichl, A.~Brandstötter, J.~Böhm, M.~Kühmayer, U.~Kuhl, and S.~Rotter,
  ``Focusing inside {Disordered} {Media} with the {Generalized}
  {Wigner}-{Smith} {Operator},'' {\em Physical Review Letters}, vol.~119, July
  2017.

\bibitem{PADiss}
P.~Ambichl, {\em \href{http://katalog.ub.tuwien.ac.at/AC13104469}{Coherent Wave
  Transport: Time Delay and Beyond}}.
\newblock PhD thesis, TU Wien, 2016.

\bibitem{fyodorov_statistics_1997}
Y.~V. Fyodorov and H.-J. Sommers, ``Statistics of resonance poles, phase shifts
  and time delays in quantum chaotic scattering: {Random} matrix approach for
  systems with broken time-reversal invariance,'' {\em Journal of Mathematical
  Physics}, vol.~38, pp.~1918--1981, Apr. 1997.

\bibitem{rotter_light_2017}
S.~Rotter and S.~Gigan, ``Light fields in complex media: {Mesoscopic}
  scattering meets wave control,'' {\em Reviews of Modern Physics}, vol.~89,
  Mar. 2017.

\bibitem{Cheng:14}
X.~Cheng and A.~Z. Genack, ``Focusing and energy deposition inside random
  media,'' {\em Opt. Lett.}, vol.~39, pp.~6324--6327, Nov 2014.

\bibitem{PhysRevLett.101.120601}
I.~M. Vellekoop and A.~P. Mosk, ``Universal optimal transmission of light
  through disordered materials,'' {\em Phys. Rev. Lett.}, vol.~101, p.~120601,
  Sep 2008.

\end{thebibliography}

\end{document}